# Artificial Intelligence Powered Identification of Potential Antidiabetic Compounds in *Ficus religiosa*

**Md Ashad Alam[1], Md Amanullah[2]**

[1]Ochsner Center for Outcomes Research, Ochsner Research, Ochsner Clinic Foundation, New Orleans, LA 70121, USA

[2]Department of Ophthalmology, Wilmer Eye Institute, Johns Hopkins University School of Medicine, Baltimore, MD 21287, USA

## Abstract

Diabetes mellitus is a chronic metabolic disorder that necessitates novel therapeutic innovations due to its gradual progression and the onset of various metabolic complications. Research indicates that *Ficus religiosa* is a conventional medicinal plant that generates bioactive phytochemicals with potential antidiabetic properties. The investigation employs ecosystem-based computational approaches utilizing artificial intelligence to investigate and evaluate compounds derived from *Ficus religiosa* that exhibit antidiabetic properties. A comprehensive computational procedure incorporated machine learning methodologies, molecular docking techniques, and ADMET prediction systems to assess phytochemical efficacy against the significant antidiabetic enzyme dipeptidyl peptidase-4 (DPP-4). DeepBindGCN and the AutoDock software facilitated the investigation of binding interactions via deep learning technology. Flavonoids and alkaloids have emerged as attractive phytochemicals due to their strong binding interactions and advantageous pharmacological effects, as indicated by the study. The introduction of AI accelerated screening procedures and enhanced accuracy rates, demonstrating its efficacy in researching plant-based antidiabetic agents. The scientific foundation now facilitates future experimental validation of natural product therapies tailored for diabetic management.

## Keywords

Artificial Intelligence; Molecular Docking Simulation; Diabetes Mellitus; Drug Screening; Phytochemicals.

## Introduction

Type 2 diabetes mellitus is a global metabolic illness characterized by persistent hyperglycemia due to impaired insulin secretion and resistance [1], [2]. Over 90 percent of diabetes mellitus cases are classified as Type 2 diabetes mellitus (T2DM), characterized by significant associations with insulin resistance and β-cell dysfunction [3], [4]. T2DM constitutes a growing global health issue, significantly burdening healthcare systems economically [5]. Research in 2021 indicated that 537 million persons globally had diabetes, with projections estimating 643 million cases by 2030 [6]. Existing antidiabetic medications necessitate the development of new therapeutic agents due to their undesirable effects, diminished efficacy, and high costs for patients [7], [8].

The therapy of diabetes emphasizes dipeptidyl peptidase-IV (DPP-IV), a serine protease enzyme recognized for its rapid breakdown of incretin hormones, such as glucagon-like peptide-1 (GLP-1) and glucose-dependent insulinotropic polypeptide (GIP) [9]. The incretin hormones enhance insulin secretion and inhibit glucagon release while regulating postprandial blood glucose levels [10]. The clinical importance of DPP-IV inhibitors, or "gliptins," is based on their superior glycemic management capabilities and a low risk of hypoglycemia [11]. The demand for secure natural substitutes for synthetic DPP-IV inhibitors is rising because of their possible harmful effects, including pancreatitis, arthralgia, and immune-related problems [12]. Multiple medical plants function as reservoirs of bioactive compounds for treating various chronic diseases, particularly diabetes [13]. People in South and Southeast Asia widely use *Ficus religiosa* (sacred fig) as an Ayurvedic medicinal plant for treating diabetes, inflammation, and other illnesses [14]. Research findings show that the hypoglycemic properties of the *Ficus religiosa* plant can be found in its bark, leaves, and fruits [15]. Studies on *Ficus religiosa* phytochemicals show the plant contains multiple secondary metabolites, including flavonoids, triterpenoids, sterols, and phenolic acids, demonstrating pharmacological effects linked to insulin regulation and glucose metabolism [16].

Limited research exists about the molecular processes through which *Ficus religiosa* phytochemicals function as antidiabetic agents, specifically through DPP-IV inhibition [17]. The existing drug discovery procedures face multiple downsides because they take excessive time, require high expenses, and show limited capacity to determine drug absorption characteristics and toxicological profiles [18]. Computational drug discovery technologies combined with artificial intelligence (AI) enable quicker exploration of new therapeutic compounds through recent technological developments [19], [20]. In silico drug screening has experienced advancements through machine learning algorithms, deep learning frameworks, and molecular modeling techniques that generate fast, accurate predictions about drug-target interactions, binding affinities, and ADMET properties [21-30].

Drug development has been transformed by deep learning, a subfield of artificial intelligence, as it offers efficient and expedited methods that decrease both time and costs relative to conventional drug discovery strategies [31-40]. Contemporary drug discovery techniques necessitate extensive time and substantial financial resources, resulting in a twelve-year timeline and expenditures amounting to several billion dollars for the development of novel pharmaceuticals [20]. Research employing deep learning methodologies mitigates drug discovery constraints by scrutinizing vast biological, chemical, and clinical datasets to produce precise insights regarding molecular characteristics, drug-target interactions, and toxicities [18]. Molecular structures and biomedical data are exceptionally analyzed using convolutional neural networks (CNNs), recurrent neural networks (RNNs), and graph neural networks (GNNs) [41]. The expedited evaluation of pharmaceutical prospects use virtual screening models in conjunction with Variational Autoencoders (VAEs) and Generative Adversarial Networks (GANs) for de novo drug design, yielding novel chemical entities that align with specified characteristics. Drug development accelerates in its initial phases due to GNN-based screening, which reduces the pool of prospective candidates prior to laboratory assessments [42].

Deep learning employs clinical trial and patient data analysis to predict drug safety levels, optimize dosages, and forecast probable adverse effects during drug development [43]. Deep learning reduces the likelihood of clinical failures in late-stage drug development, which are the most costly and perilous phases in pharmaceutical progress. The technique engages in medication repurposing efforts by identifying novel therapeutic uses for existing pharmaceuticals. Efforts to revitalize drugs garnered significant attention during the COVID-19 pandemic [44], [45]. The extensive advantages of deep learning implementation in drug development encounter challenges such as the inability to obtain high-quality data, complexities in model interpretation, and the necessity for diversified professional collaboration. Deep learning is expected to enhance its significant role in advancing precision medicine by integrating with genetic research and biological systems.

The research employed AI-driven systems to determine whether phytochemicals from *Ficus religiosa* exhibit antidiabetic effects by inhibiting DPP-IV activity. The analysis commenced with developing a structurally optimized DPP-IV protein target, accompanied by a selection of bioactive chemicals sourced from *Ficus religiosa*. In this study, we conducted energy minimization on ligand structures before processing proteins and ligand data files to assess molecular docking and virtual screening [41]. The DeepBindGCN model operated as a deep learning screening technique utilizing graph convolutional networks to enhance estimates of ligand binding probabilities [23]. ADMET profiling of the tested drugs was conducted using the ADMETlab 2.0 platform, facilitating the evaluation of pharmacokinetic characteristics and toxicity potential [46]. The chosen ligands exhibited their binding affinity to DPP-IV via molecular docking with AutoDock Vina, followed by MM/PBSA calculations to evaluate significant interactions with the target. This study integrates traditional knowledge of medicinal plants with advanced computational systems to identify lead compounds for developing contemporary antidiabetic drugs, including safe and effective natural DPP-IV inhibitors. This research enhances understanding of the biological activities of *Ficus religiosa* through molecular insights and illustrates how artificial intelligence revolutionizes natural medication assessments.

## Methods

### Protein target preparation

The human dipeptidyl peptidase IV (DPP-IV) is the protein target due to its significant function in regulating glucose metabolism and clinical relevance in treating type 2 diabetes mellitus [47]. The crystal structure of DPP-IV is obtained from the Protein Data Bank, identified by PDB ID: 1J2E [48]. This structural information presents an inhibitor-bound human dipeptidyl peptidase IV enzyme and comprehensive details regarding its requisite binding interactions and active site constituents, and the protein required some preparation via PyMOL and AutoDock Tools before employing virtual screening and molecular docking analyses [49], [50]. The computational model necessitated the elimination of water molecules, non-standard amino acid residues, and co-crystallized ligands to guarantee unobstructed docking procedures. The protein structure's protonation entails adding crucial hydrogen atoms, facilitating accurate hydrogen bond interaction modeling. The protein atoms were assigned computed Gasteiger partial charges, facilitating accurate docking calculations. The protein structure was prepared for workflows after cleaning, formatting, and subsequent saving in PDBQT format. During preprocessing, the docking results gain reliability as the approach accurately represents the binding circumstances within the enzyme's active site.

### Ligand preparation

Researchers compiled a library of bioactive chemicals derived from *Ficus religiosa*, a historically utilized antidiabetic medicinal plant. A comprehensive review procedure and database searches of IMPPAT [51], PubChem [52], and ChEBI [51] discovered the phytochemical constituents in *Ficus religiosa*. A computational algorithm has selected chemicals from *Ficus religiosa* that exhibit potential as active antidiabetic medicines. The 2D structures from PubChem were transformed into 3D structures via Open Babel and retrieved in SDF format [53]. The compounds were subjected to MMFF94 force field energy minimization, resulting in stable conformations that enhance the accuracy of docking data. The PDB format containing 3D structures was subjected to further processing with AutoDock Tools after completing minimization procedures[54]. The preprocessing phase involved delineating torsional flexibility, amalgamating non-polar hydrogens, and computing Gasteiger partial charges for each ligand. The PDBQT format converted each ligand before its preservation for AutoDock Vina compatibility. This preprocessing method optimized the structure of *Ficus religiosa* phytochemical compounds and formatted them appropriately for molecular docking to evaluate their potential inhibitory effects on the DPP-IV target.

**Artificial intelligence-based ADMET prediction**

The assessment of *Ficus religiosa* phytochemicals concentrated on ADMET predictions utilizing ADMETlab 2.0, accessible at https://admet.ai.greenstonebio.com/ [46]. The two-dimensional chemical structures of the chosen compounds were acquired either in SMILES format from the PubChem database or produced via the integration of ChemDraw and Open Babel tools. ADMETlab 2.0 accepted the supplied SMILES strings to produce predictions via machine learning models utilizing well-documented experimental datasets.

ADMETlab 2.0 functions as a sophisticated AI platform that predicts essential drug-related features, including Absorption, Distribution, Metabolism, Excretion, and Toxicity (ADMET) of chemical substances. The successful advancement of drug candidates is heavily contingent upon these features, as inadequate ADMET profiles are a primary cause of failures in late-stage drug development. The second iteration of ADMETlab augments its capabilities by integrating advanced machine learning techniques, expanded data repositories, and a more user-friendly interface. ADMETlab 2.0 produces approximately 300 predictions on drug-likeness, encompassing assessments of blood-brain barrier permeability, oral bioavailability, cytochrome P450 interactions, as well as evaluations of hepatotoxicity and cardiotoxicity. ADMETlab 2.0 attains accurate predictions by employing neural networks, support vector machines, and ensemble learning AI models, while the identification of potential dangers commences early as researchers utilize ADMETlab to discover optimal therapeutic substances.

ADMETlab 2.0 offers a significant advantage by diminishing the duration required for drug discovery operations. ADMET property evaluation necessitated laborious and costly in vitro and in vivo experimental testing using conventional methodologies. The AI prediction tool ADMETLab 2.0 allows researchers to analyze large compound libraries for unfavorable characteristics, hence conserving resources for experimental testing on viable candidates. Users of this tool can execute iterative alterations on chemical structures to obtain immediate feedback on their impact on ADMET properties [46]. AI technology facilitates real-time structure-property feedback, assisting medication designers in optimizing leads. ADMETlab 2.0 demonstrates how artificial intelligence transforms pharmaceutical drug research in its early phases, improving productivity while increasing medicine success rates and reducing resource consumption and development time. AI-driven platforms will play a crucial role in expediting the discovery of safer and more effective therapeutics due to advancements in AI technology [43].

The analysis identified the outcomes for human intestinal absorption (HIA) penetration, blood-brain barrier (BBB) penetration, plasma protein binding (PPB), and cytochrome P450 enzyme interactions as pharmacokinetic parameters. The toxicity assessments examined Ames mutagenicity, hERG inhibition, hepatotoxicity, and LD50 [43]. The drug-like qualities of each chemical were assessed using three primary criteria: Lipinski's Rule of Five, Veber's Rule, and medicinal chemistry filtration. The ADMET profiles of promising compounds qualified them for progression into possible DPP-IV inhibitor development. ADMET prediction was important in eliminating drugs with inferior pharmacokinetic properties or heightened toxicity risks, assisting researchers in identifying effective and safe lead candidates.

**Deep learning-based ligand screening**

Virtual screening enhanced its accuracy and efficiency by utilizing DeepBindGCN, an improved ligand screening technique based on graph convolutional networks (GCN) [23]. The study used PDB ID: 1J2E for human dipeptidyl peptidase IV (DPP-IV) as the principal three-dimensional target structure. The entity associated with the ligand used for crystallization provided the foundation for generating an appropriate format compatible with DeepBindGCN [55]. The development team processed the phytochemicals of *Ficus religiosa* before their preparation for use as ligands. The chemical structures were obtained from PubChem and additionally extracted manual structures, which they transformed into nodes and edges to represent these compounds as graph-based molecular representations. The graph-based inputs functioned as essential components that allowed DeepBindGCN to comprehend the spatial and chemical factors.

DeepBindGCN is a sophisticated deep learning framework that utilizes graph neural networks (GNNs) to predict drug-target binding affinity, hence facilitating the early stages of drug discovery [23]. Anticipating the affinity of pharmacological molecules for target proteins is crucial for enhancing the efficacy of novel therapeutic development, as it reduces both production timelines and costs. DeepBindGCN utilizes graph-based methodologies to analyze molecules and proteins using node-and-edge representations of atoms and amino acids, facilitating enhanced detection of molecule-protein interactions. The model advantages itself from its graph-based data structure, resulting in enhanced processing of intricate relationships within the data. Two distinct GNN encoders progressively evaluate drug and protein graphs, subsequently merging their extracted features to assess binding affinity levels. DeepBindGCN use Graph Neural Networks to analyze intricate molecular data for predicting drug-target interactions in novel combinations.

The primary benefit of DeepBindGCN is in its integrated learning framework, which combines feature extraction with predictive tasks without necessitating manual feature input. DeepBindGCN outperforms traditional machine learning methods and other deep learning techniques on the Davis and KIBA benchmark datasets due to its design. The interpretative abilities of DeepBindGCN are facilitated by its attention mechanisms, which identify critical substructures inside molecules that significantly influence binding. The system offers critical information that enables scientists to corroborate trial outcomes and enhance rational drug development techniques. DeepBindGCN models now expedite the drug development process due to the rapid accumulation of biological and chemical data. The amalgamation of efficiency, interpretability, and accuracy positions DeepBindGCN as a significant computational tool for drug discovery and precision medicine.

The DeepBindGCN preprocessing pipeline processed protein and ligand files to generate protein-ligand complex graphs. Upon training the DeepBindGCN model, it acquired the ability to ascertain binding scores indicating the likelihood and strength of ligand binding to DPP-IV. Variations in binding scores from ligands facilitated the evaluation of potential DPP-IV inhibition candidates. The AI-driven system developed an efficient approach to finding interesting bioactive chemicals by evaluating binding potential, reducing the need for conventional docking methods.

**Molecular Docking**

AutoDock Vina conducted docking simulations assessing ligand-DPP-IV binding energies and their interaction patterns [56]. The docking grid positioned its center on the active site residues identified in the co-crystallized ligand binding region of the 1J2E structure, simultaneously encompassing the substrate binding area. The approach enhanced grid box diameters and sample thoroughness to attain accurate sampling outcomes. The ranking method evaluated compounds based on their binding affinity, quantified in kcal/mol to advance further inquiry. The assessment of premier ligands via visualization and analysis was conducted using PyMOL [57]. The evaluation of inhibitory potential necessitated the examination of significant connections among essential residues by evaluating hydrogen bonds, hydrophobic contacts, and π-π stacking interactions. The design method prioritized ligands that exhibited binding patterns like the natural inhibitor.

**Binding free energy calculations**

A more accurate assessment of the binding of *Ficus religiosa* phytochemicals to the DPP-IV target protein necessitated Molecular Mechanics/Poisson–Boltzmann Surface Area (MM/PBSA) simulations [58]. This approach facilitates binding free energy estimations by evaluating docked ligand-protein complexes by integrating molecular mechanics energies with surface area and solvation model components. Autodock Vina produced the most optimal docked complexes, which were further examined for additional investigation.

**Results**

**Protein and ligands**

A research study illustrated the molecular docking methodology for discovering human dipeptidyl peptidase IV (DPP-IV) inhibitors, which serve as a crucial enzyme in glucose metabolism for type 2 diabetes mellitus [12]. Fig 1 illustrates a multicolored ribbon model on its left side, representing the DPP-IV protein structure, including its structural domains, α-helices, and β-sheets as secondary structures. A red-highlighted region denotes the enzyme's docking location, where ligands underwent computational assessments of binding interactions. Fig 1 displays twelve phytochemicals derived from *Ficus religiosa*, illustrated by their corresponding 2D chemical molecular representations. Compounds derived from *Ficus religiosa* encompass bergaptol and bergapten from coumarins, methyl oleanolate, lupeol, and lupenone from triterpenoids, beta-sitosterol, lanosterol, and stigmasterol from steroids, caffeic acid from phenolics, and beta-sitosterol-d-glucoside from glycosides. The chemical diversity of these compounds arises from their backbone architectures, which incorporate both hard aromatic rings and flexible long-chain alcohols, together with complex steroidal configurations that exhibit various binding modes to the DPP-IV active site.

The hydroxyl, methoxy, and carbonyl functional groups of substances facilitate the formation of hydrogen bonds or hydrophobic interactions with the residues of the DPP-IV binding pocket. Caffeic acid and bergenin have several hydroxyl groups that can form robust polar bonds. In contrast, lupeol and stigmasterol create hydrophobic interactions and van der Waals forces within the enzyme's active site. The compounds' structural variety enhances the possibility of identifying multiple binding configurations and inhibition strategies during virtual screening.

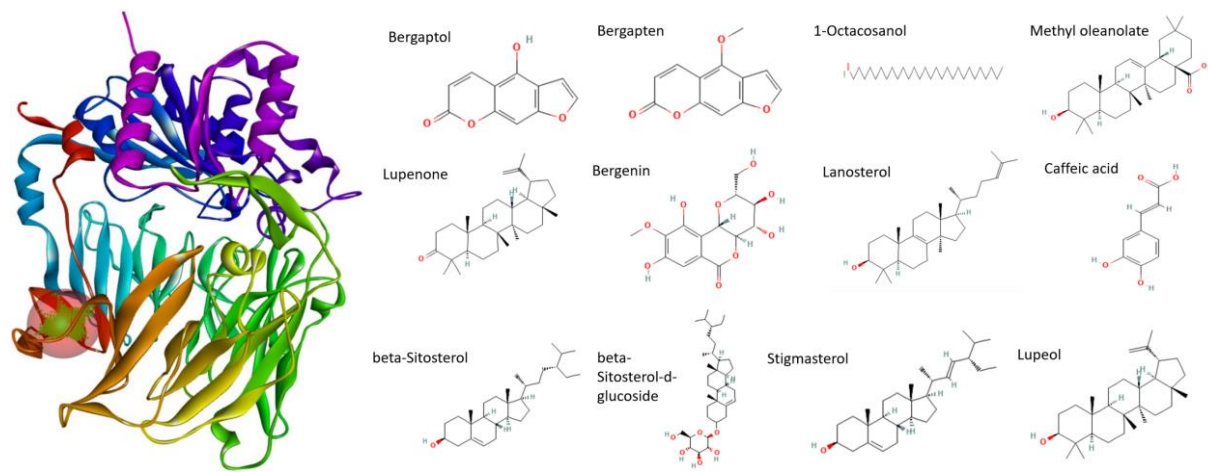

**Fig 1**. Molecular structure of human dipeptidyl peptidase IV (DPP-IV) shown with its active site highlighted (left), alongside 2D structures of selected phytochemicals from *Ficus religiosa* (right).

## ADMET properties

Table 1 delineates the critical characteristics of *Ficus religiosa* phytochemicals and concerns about drug-likeness and oral bioavailability. The phytochemical compounds have molecular weights ranging from 180.19 g/mol for caffeic acid to 576.85 g/mol for beta-sitosterol-d-glucoside. The beta-sitosterol-d-glucoside component marginally exceeds Lipinski's Rule's 500 g/mol limit, facilitating oral bioavailability, as its molecular weight is 576.85 g/mol. The lipophilicity of the compounds in this study is quantified by LogP values ranging from the highly hydrophilic -1.20 for bergenin to the highly hydrophobic 10.14 for 1-octacosanol. The solubility and accessibility of substances diminish as their logP values exceed five due to the emergence of pronounced lipophilic properties. The dataset indicates that stigmasterol, beta-sitosterol, lupeol, lupenone, and 1-octacosanol exhibit high logP values above 7, as negative solubility scores corroborate their low solubility.

Properties not pertain to hydrogen bond acceptors or donors should be disregarded when evaluating permeability and solubility. The pronounced polar characteristics of Bergenin, with nine acceptors and five donors, and beta-sitosterol-d-glucoside, with six acceptors and four donors, are evidenced by their TPSA values of 145.91 Å² and 99.38 Å², respectively. Membranes exhibit restricted permeability when they incorporate substances with elevated TPSA values. The passive membrane transport potential of stigmasterol and lupenone molecules is expected to enhance, given their TPSA values below 20 Å². Most compounds adhere to Lipinski's criteria by achieving fewer than four infractions. Bioavailability concerns may arise for beta-sitosterol-d-glucoside and 1-octacosanol, as seen by their violation outcomes. Poor water solubility adversely impacts most substances since their solubility values generally yield adverse effects, particularly among triterpenoids and sterols like lupeol and stigmasterol. The superior bioavailability and elevated solubility score of caffeic acid (21.48%) render this compound a more advantageous option for oral pharmaceutical development.

**20Table 1**. Physicochemical and pharmacokinetic properties of selected Ficus religiosa phytochemicals, including molecular weight, lipophilicity (logP), hydrogen bond acceptors/donors, Lipinski's rule compliance, topological polar surface area (TPSA), solubility, and predicted bioavailability.

| Ligand | Molecular Weight | logP | Hb acceptors | Hb donors | Lipinski | TPSA | Solubility | Bioavailability |
|---|---|---|---|---|---|---|---|---|
| Bergenin | 328.27 | -1.20 | 9 | 5 | 4 | 145.91 | -1.47 | 13.10 |
| Caffeic acid | 180.19 | 1.19 | 3 | 3 | 4 | 77.76 | -1.64 | 21.48 |
| Bergapten | 216.19 | 2.54 | 4 | 0 | 4 | 52.58 | -3.68 | 50.21 |
| Bergaptol | 202.16 | 2.24 | 4 | 1 | 4 | 63.58 | -3.43 | 18.84 |
| beta-Sitosterol-d-glucoside | 576.85 | 5.84 | 6 | 4 | 2 | 99.38 | -6.42 | 18.22 |
| Stigmasterol | 412.70 | 7.80 | 1 | 1 | 3 | 20.23 | -6.76 | 32.41 |
| beta-Sitosterol | 414.71 | 8.02 | 1 | 1 | 3 | 20.23 | -6.65 | 31.05 |
| Lupeol | 426.72 | 8.02 | 1 | 1 | 3 | 20.23 | -7.21 | 28.38 |
| Lupenone | 424.71 | 8.23 | 1 | 0 | 3 | 17.07 | -6.57 | 42.34 |
| 1-Octacosanol | 410.77 | 10.14 | 1 | 1 | 3 | 20.23 | -6.62 | 11.98 |

Fig 2 illustrates a comparative analysis of DrugBank reference compounds, represented as blue circles, alongside input molecules, depicted as red stars, across four pharmacokinetic and toxicity evaluations: human intestinal absorption (A), blood-brain barrier penetration (B), cell permeability (C), and carcinogenicity (D). All plots employ projected clinical toxicity probability on the y-axis, exhibiting diverse distribution through marginal histograms.

Panel A data indicates that input compounds exhibit high human intestinal absorption (HIA) predictions (exceeding 0.8 likelihood) and low clinical toxicities (below 0.2). The amalgamation of efficient oral absorption properties with these safety attributes indicates favorable oral bioavailability. Panel B's evaluation indicates that the input compounds exhibit moderate to high permeability across the blood-brain barrier while maintaining acceptable toxicity levels, suggesting possible advantages for the central nervous system without associated safety risks. The aggregation of input molecules exhibits elevated cell effective permeability values (log scale) alongside negligible toxicity levels in panel C. Panel D demonstrates that the input compounds exhibit an exceedingly low probability of carcinogenicity, suggesting their therapeutic safety potential in treatment applications. The input compounds exhibit positioning within advantageous regions of the pharmacokinetic-toxicity spectrum that align with and exceed numerous reference compounds listed in DrugBank. The encouraging results suggest that the medicines are promising for future development, as they demonstrate elevated absorption rates, enhanced permeability characteristics, and minimal safety hazards.

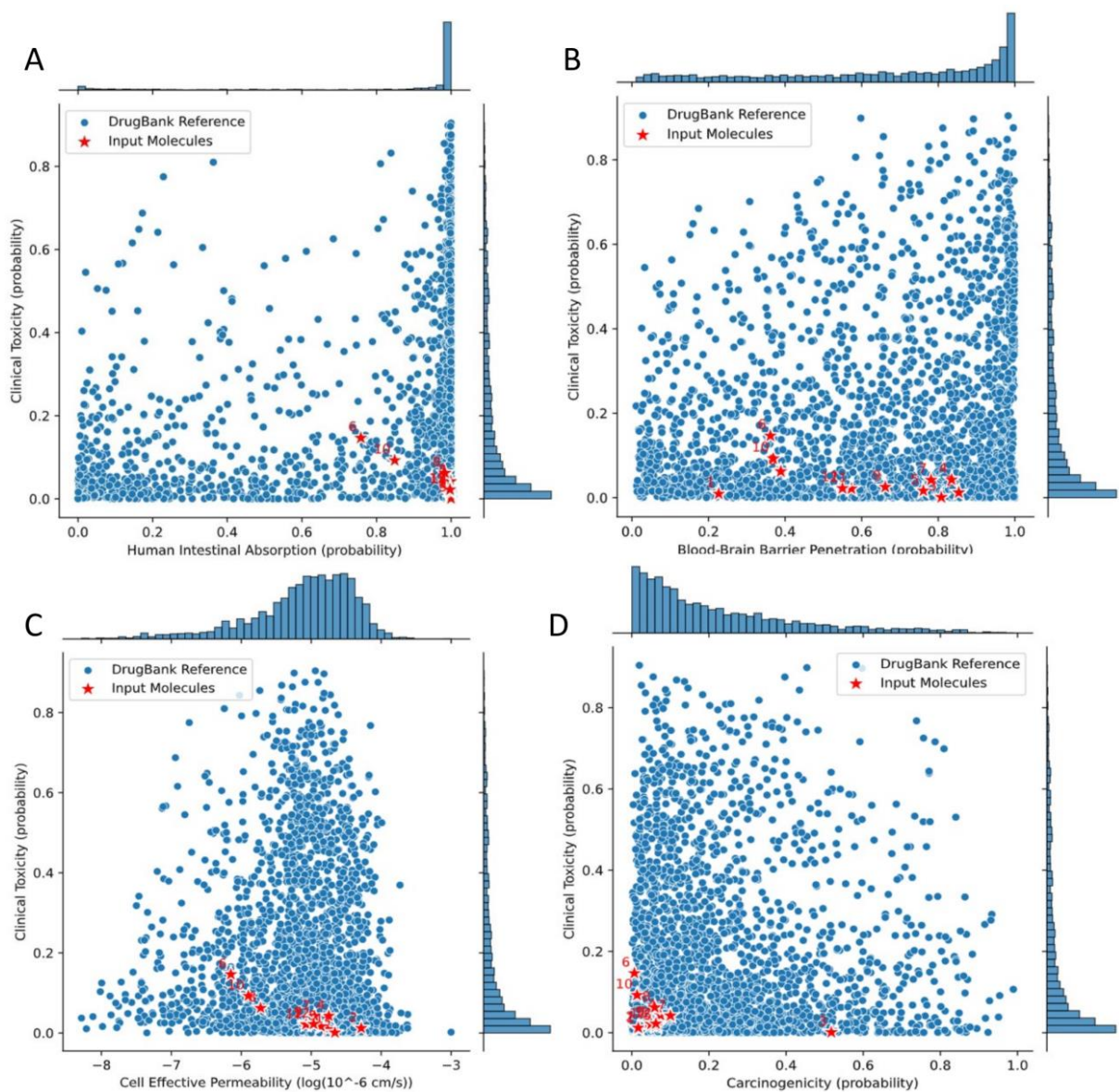

**Fig 2**. Comparative scatter plots of input molecules (red stars) versus DrugBank reference compounds (blue circles) across four pharmacokinetic and toxicity parameters. (A) Human intestinal absorption, (B) Blood-brain barrier penetration, (C) Cell effective permeability, and (D) Carcinogenicity, all plotted against clinical toxicity probability. Marginal histograms

show the distribution of each parameter. Input molecules consistently exhibit low toxicity and favorable ADMET properties.

**Protein-ligand interactions**

The data presented in Table 2 includes docking scores and DeepBindGCN machine-learning predictions concerning binding affinities for ligands evaluated on BC and RG proteins. BC denotes a likely bile duct cancer target protein, and RG is an alternate reference protein. The docking scores quantify the expected energy levels of protein-ligand interactions, with lower binding energies indicated by more negative values. DeepBindGCN predictions indicate the binding affinity between ligands and BC and RG, with higher values corresponding to stronger anticipated interactions.

**Table 2**. Docking scores and DeepBindGCN-predicted binding affinities of selected ligands against the BC target and RG reference. Lower docking scores indicate stronger binding, while higher DeepBindGCN values reflect greater predicted binding affinity.

| Ligand | Docking score | DeepBindGCN_BC | DeepBindGCN_RG |
|---|---|---|---|
| Bergenin | -6.621 | 1 | 6.079 |
| Caffeic acid | -6.589 | 1 | 6.025 |
| Bergapten | -4.028 | 1 | 4.666 |
| Bergaptol | -3.479 | 1 | 4.041 |
| beta-Sitosterol-d-glucoside | -3.333 | 1 | 3.036 |
| Stigmasterol | -2.917 | 0 | 2.116 |
| beta-Sitosterol | -2.825 | 0 | 2.022 |
| Lupeol | -2.801 | 0 | 2.012 |
| Lupenone | -2.374 | 0 | 1.996 |
| 1-Octacosanol | -0.941 | 0 | 0.448 |

The docking data reveal that bergenin and caffeic acid are the leading compounds, exhibiting docking scores of -6.621 and -6.589, respectively, coupled with identical DeepBindGCN scores of 1 for BC and scores of 6.079 and 6.025 for RG. Concurrent validation by docking and machine learning yields dependable outcomes that indicate robust expected binding affinity in attractive candidates for further evaluation. The docked scores of bergapten and bergaptol fall within an intermediate range, while their DeepBindGCN_RG readings exhibit average values between 4 and 4.6; nonetheless, they show complete binding to BC (DeepBindGCN_BC = 1).

Phytosterols, including beta-sitosterol-d-glucoside, stigmasterol, beta-sitosterol, and lupeol, exhibit docking scores ranging from -3.3 to -2.8, whereas their DeepBindGCN_RG predictions vary between 1.996 and 3.036. The docked compounds have diminished predicted binding affinity for BC (DeepBindGCN_BC = 0), except beta-sitosterol-d-glucoside. Consequently, they demonstrate less target-specific interactions. Based on the computational study, Bergenin, and caffeic acid exhibit strong binding properties, establishing them as optimal selections.

Two ligands engage with a target protein via molecular docking, as illustrated by the accompanying 3D structural models and 2D interactive mappings in Fig 3. The three-dimensional configurations of protein-ligand complexes are depicted in both Panels A and B. In addition to the protein backbone represented as ribbons, which include cyan β-sheets, red α-helices, and green loop structures, the ligands are depicted using stick models to illustrate their placement within the binding pockets. The ligands reside comfortably in a binding site between different secondary structures, signifying a stable binding environment. The two comprehensive 2D interaction diagrams depicted in panels C and D delineate particular amino acid interactions and diverse interaction types of the bound ligands. The ligand forms several hydrogen bonds and van der Waals interactions with essential residues LYS71, GLU73, and SER59. The binding interface, governed by hydrogen bonding and polar interactions, guarantees precise ligand interaction and stable binding conditions. The electrostatic interactions between the second ligand and LYS56 and ARG54, along with hydrogen bonds with LYS71 and SER59, are illustrated in the panels of D. The binding interaction assumes a more electric character.

The sequentially interacting ligands with the protein structure involve many intermolecular forces, notably van der Waals forces, hydrogen bonds, and ionic interactions, which collectively exhibit a robust binding affinity. The evolving dynamics of interatomic interactions between ligands and receptors elucidate distinctive binding characteristics that modify functional effects and selectivity towards targets. The research validates the structural findings of the docking model and offers critical insights to improve ligand interactions for enhanced binding potential.

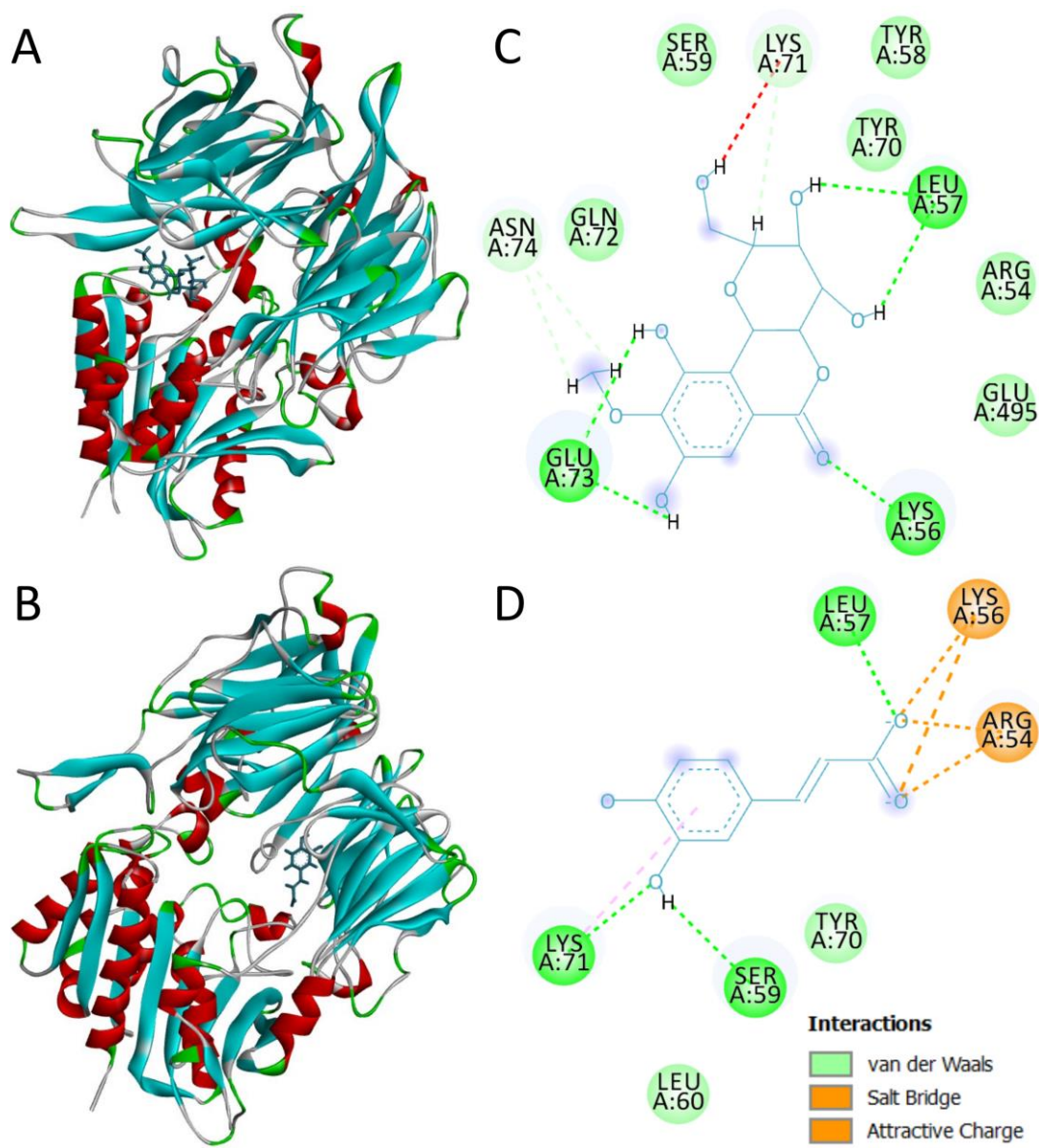

**Fig 3**. Structural and interaction analysis of two ligand-protein complexes. (A, B) 3D ribbon representations showing ligand binding within the protein structure. (C, D) 2D interaction diagrams illustrating hydrogen bonds, van der Waals forces, salt bridges, and electrostatic interactions between ligands and key amino acid residues.

## Binding free energy of complexes

Table 3 comprehensively analyzes the energies associated with ligand-protein interactions derived from molecular docking experiments. The study presents data on van der Waals energy (even), electrostatic energy (could), docking score or binding energy model (model), overall binding energy, and internal strain or torsional energy (internal) for each ligand. The parameters provide substantial insights into ligand-binding interactions' thermodynamic favorability and structural robustness. Bergenin demonstrates strong ligand-protein interactions, evidenced by its low model value of -52.09, whereas the cumulative review and could energy measurements total -39.73 kcal/mol. The conformational strain during binding is minimal, as indicated by an internal score of 5.42. Caffeic acid demonstrates a favorable binding energy of -23.71 kcal/mol, resulting from a harmonious interplay of electrostatic and van der Waals interactions alongside constrained internal energy. The docking results previously indicated that these drugs have robust binding capabilities according to the new experimental data.

**Table 3**. The table presents the van der Waals energy (evdw), electrostatic energy (ecoul), docking score or binding energy model (emodel), overall binding energy, and internal strain or torsional energy (einternal) for each ligand.

| Ligand | evdw | ecoul | emodel | energy | einternal |
|--------|------|-------|--------|--------|-----------|
| Bergenin | -15.49 | -24.24 | -52.09 | -39.74 | 5.42 |
| Caffeic acid | -13.57 | -10.14 | -34.69 | -23.71 | 2.89 |
| Bergapten | -19.19 | -4.47 | -29.39 | -23.67 | 0.06 |
| Bergaptol | -15.51 | -3.32 | -27.71 | -18.84 | 0.00 |
| beta-Sitosterol-d-glucoside | -28.20 | -14.64 | -49.99 | -42.85 | 5.69 |
| Stigmasterol | -18.94 | -6.79 | -29.59 | -25.73 | 2.17 |
| beta-Sitosterol | -19.68 | -6.24 | -29.75 | -25.92 | 1.96 |
| Lupeol | -26.28 | -3.31 | -34.40 | -29.60 | 0.97 |
| Lupenone | -28.26 | 0.54 | -30.16 | -27.72 | 2.68 |
| 1-Octacosanol | -27.10 | -2.65 | -28.28 | -29.75 | 6.95 |

Three sterol derivatives, namely beta-sitosterol-d-glucoside, stigmasterol, and lupeol, benefit from significant van der Waals interactions (evdw up to -28.26) due to their considerable hydrophobic structural components; however, these compounds demonstrate elevated internal energy, indicating conformational strain. The half-browser binding energy of beta-sitosterol-d-glucoside at -42.85 impedes its conformational stability owing to its high internal value of 5.69. Despite 1-octacosanol and lupenone exhibiting significant van der Waals contributions (-27 to -28), their inadequate electrostatic binding interactions, along with elevated internal energies (6.95 and 2.68), appear to limit their effective binding potential in biological systems.

**Discussion**

The integration of Artificial Intelligence (AI) in biomedical research is on the rise, as it offers sophisticated tools that accelerate the discovery of medicinal molecules from natural sources [59]. The examination of plant-derived bioactive chemicals by AI methodologies effectively identifies their antidiabetic effects. Individuals worldwide confront type 2 diabetes mellitus, a health issue characterized by insulin resistance and impaired glucose metabolism [60].

Traditional medicine employs plants as an approach for diabetes control, as these plants encompass four types of chemicals that exhibit potential antidiabetic characteristics [61]. The chemicals comprise flavonoids, alkaloids, terpenoids, and polyphenols. Conventional laboratory assays for verifying these substances are time-consuming and costly to execute [26]. AI provides a supplementary approach that accelerates analytical processes through data-driven predictions and pattern recognition.

The integration of AI methodologies, including machine learning and deep learning, allows analysts to evaluate complex biological data, leading to predictions about the pharmacological properties of plant-derived substances [62]. The primary phase necessitates researchers to obtain data from databases that encompass information on plant metabolites, chemical structures, and their recorded pharmacological effects. AI algorithms utilize these datasets to learn and identify correlations between chemical structures and biological activities. The evaluation of substances for their enzyme inhibition potential in glucose metabolism is conducted using support vector machines (SVM), random forests, and neural networks [63]. The Artificial Intelligence system also offers predictions regarding the interactions between plant components and insulin receptors, as well as essential proteins that regulate blood sugar levels.

Virtual screening is a crucial application as it use AI models to forecast the chemical binding interactions between phytochemicals and diabetic target receptors. AI is essential for identifying a limited selection of natural chemicals from vast collections of substances for subsequent experimental laboratory assessments [64]. AI employs molecular docking and dynamics simulations to furnish comprehensive insights into compound-receptor or compound-enzyme molecular interactions, thereby assisting researchers in identifying compounds with the most promising inhibitory or activating characteristics [24]. The utilization of AI methodologies facilitates network pharmacological analyses for diabetes, enabling researchers to examine its intricate multifactorial attributes. These approaches replicate the interactive mechanisms between plant-derived bioactive compounds and several targets and pathways concurrently, as phytochemicals generally demonstrate systemic effects. The processing of genomic, proteomic, and metabolomic data using artificial intelligence aids researchers in comprehending the systemic biological effects of plant substances with antidiabetic capabilities [65]. The investigation of the antidiabetic characteristics of bioactive plant components employs artificial intelligence as a transformative instrument in contemporary scientific methodology. AI technology expedites the discovery of natural therapeutic items by optimizing the processes of identification, prediction, and validation, yielding superior outcomes without extensive laboratory experimentation. Improved accessibility to high-quality biological and chemical data enhances the predictive capabilities of AI models, facilitating the development of safer plant-derived therapies for diabetes.

Researchers used artificial intelligence to examine the antidiabetic properties of bioactive chemicals from *Ficus religiosa* and assess their interaction with human dipeptidyl peptidase IV (DPP-IV), a recognized target enzyme for type 2 diabetes mellitus [66]. Prescreens of bioactive compounds that incorporated ADMET prediction by machine learning, deep learning ligand screening, and binding free energy calculations established an efficient multi-tiered computational candidate selection methodology [67]. The discovered phytochemical compounds exhibit promising traits for prospective use as natural DPP-IV inhibitor leads in pharmaceutical research.

Among the chemicals isolated from *Ficus religiosa*, bergenin, and caffeic acid had the highest affinity in molecular docking studies. The interaction between bergenin and caffeic acid with three critical active site residues (LYS71, SER59, and GLU73) of DPP-IV is characterized by persistent binding structures resulting from many hydrogen bonds and van der Waals interactions. The anticipated binding scores derived from DeepBindGCN corroborated these research findings, revealing high bergenin and caffeic acid scores. Predictive systems employed these ligands with constant efficacy, indicating effective and selective target engagement. Many hydroxyl groups and polar functions in bergenin and caffeic acid molecules facilitate robust electrostatic and hydrogen-bonding interactions, enhancing their inhibitory effects on DPP-IV.

Robust van der Waals interactions were seen between triterpenoids and sterols such as lupeol, stigmasterol, and beta-sitosterol; nevertheless, these compounds exhibited low to moderate binding scores owing to their considerable hydrophobic surface area. The elevated internal energy of these compounds suggested that they underwent structural deformations upon binding to the protein. The conformational rigidity of these molecules presumably inhibits them from achieving optimal orientation within the DPP-IV active site. The DeepBindGCN study indicated that the substantial chemical entities earned diminished scores from the protein due to their relatively inadequate adaptation to the structural restrictions within the binding pocket. The compounds demonstrate advantageous lipophilicity due to elevated logP values; nevertheless, their low aqueous solubility and high logP introduce ambiguities regarding medication absorption and pharmacological efficacy.

The ADMET profiling screening findings assessed the ligands' potential utility for pharmaceutical development. The oral bioavailability of most substances, including bergenin and caffeic acid, demonstrated significant promise with few toxicity risks. The elevated water solubility and superior intestinal absorption render caffeic acid appropriate for clinical application as an oral medication formulation. The drug-likeness criteria and the low solubility characteristics of beta-sitosterol-d-glucoside and 1-octacosanol impede their therapeutic efficacy despite favorable docking scores. Testing revealed that the majority of the phytochemicals adhered to Lipinski's and Veber's drug-like criteria. Pharmacokinetic and toxicity scatterplot results demonstrated that the primary ligands exhibited minimal carcinogenic risk, elevated permeability, and reduced clinical toxicity.

MM/PBSA binding free energy calculations indicated that bergenin and caffeic acid are thermodynamically stable binding partners for DPP-IV. Bergenin attained optimal total binding energy due to its advantageous electrostatic and van der Waals interactions and little torsional strain. The sterol derivatives exhibited significant van der Waals interactions; nevertheless, their elevated internal energy and weak electrostatic forces constrained their binding capacity. These findings indicate that stable protein-ligand complexes depend on an appropriate equilibrium of intermolecular forces and minimum structural penalties.

**Conclusion**

This scientific study indicates that bergenin and caffeic acid, derived from Ficus religiosa, have notable dipeptidyl peptidase-IV (DPP-IV) inhibitory capabilities that facilitate the breakdown of incretin hormones involved in the regulation of blood glucose levels. The compounds show potential for antidiabetic medication development as they inhibit DPP-IV activity, resulting in increased insulin secretion and enhanced glycemic regulation. Researchers utilized a combination of artificial intelligence screening techniques to identify these chemicals through molecular docking and predictive modeling methodologies. The robust binding associations between bergenin and caffeic acid with the DPP-IV active site suggest their potential as effective inhibitors. The utilization of artificial intelligence in screening expedited the evaluation of natural compounds by enabling rapid, targeted testing of extensive libraries, thereby eliminating costly preliminary laboratory assessments.

Further experimental testing is required to validate the promising in silico data concerning the DPP-IV inhibitory activities of bergenin and caffeic acid before confirming medical safety and pharmacological functionalities. Laboratory studies and molecular dynamics simulations must assess bergenin and caffeic acid compounds via enzyme inhibition assays, cellular experiments, and in vivo testing to ascertain their stability, therapeutic levels, and bioavailability under physiological conditions. Subsequent research inquiries will determine whether these substances may advance as therapeutic interventions for diabetes. The study illustrates how artificial intelligence improves the examination of natural products as a means to develop novel prospective medication candidates for conventional medicinal uses. The AI-assisted approach establishes a reliable system that effectively identifies and evaluates bioactive compounds from plant sources, resulting in the advancement of safer diabetes treatment alternatives.